\documentclass[12pt,graphics]{iopart}

\newcommand{\Journal}[4]{{#1} #4 {\bf #2}, #3 }
\usepackage{iopams}  
\usepackage{graphicx}
\usepackage{epsfig}
\usepackage{units}
\usepackage{multicol,graphics,verbatim}
\usepackage{longtable}
\usepackage{colortbl}
\usepackage{color}
\usepackage{rotating}  
\usepackage{calc} 
\usepackage{upgreek}          
\usepackage{ifthen}
\usepackage[colorinlistoftodos]{todonotes}
\usepackage{lineno}

\usepackage{hyperref}

\usepackage{threeparttable}

\newcommand{\ra}{\rightarrow }
\def\Journal#1#2#3#4{{#1} {\bf #2}, #3 (#4)}

\def\ARI{\em Appl. Radiat. Isot.}

\def\NIMA{{\em Nucl. Instrum. Methods} A}

\def\NPA{{\em Nucl. Phys.} A}

\def\JPG{\em Journal of Physics G}

\def\PRC{{\em Phys. Rev.} C}

\def\EPA{{\em Europ. Phys. J.} A}

\def\EPL{\em Europhys. Lett.} 

\newcommand{\be}{\begin{equation}}
\newcommand{\ee}{\end{equation}}
\def\bea{\begin{eqnarray}} 
\def\eea{\end{eqnarray}} 
 
\newcommand{\zbb}{2\mbox{$\nu\beta\beta$ - decay} }

\newcommand{\ema}{\mbox{$\langle m_{\nu_e} \rangle $}} 
 
\newcommand{\bnel}{\mbox{$\bar{\nu}_e$} }

\begin{document}
\definecolor{brown}{RGB}{139,69,19}

\newcommand{\nuc}[2]{$^{#2}\rm #1$}

\newcommand{\bb}[1]{$\rm #1\nu \beta \beta$}
\newcommand{\bbm}[1]{$\rm #1\nu \beta^- \beta^-$}
\newcommand{\bbp}[1]{$\rm #1\nu \beta^+ \beta^+$}
\newcommand{\bbe}[1]{$\rm #1\nu \rm ECEC$}
\newcommand{\bbep}[1]{$\rm #1\nu \rm EC \beta^+$}

\newcommand{\largeGERDA}{{LArGe}}
\newcommand{\PI}{\mbox{\textsc{Phase\,I}}}
\newcommand{\PIa}{\mbox{\textsc{Phase\,Ia}}}
\newcommand{\PIb}{\mbox{\textsc{Phase\,Ib}}}
\newcommand{\PIc}{\mbox{\textsc{Phase\,Ic}}}
\newcommand{\PII}{\mbox{\textsc{Phase\,II}}}

\newcommand{\nPlus}{\mbox{n$^+$ electrode}}
\newcommand{\pPlus}{\mbox{p$^+$ electrode}}

\newcommand{\aoe}{$A/E$}

\newcommand{\order}[1]{\mbox{$\mathcal{O}$(#1)}}

\newcommand{\pic}[5]{
       \begin{figure}[ht]
       \begin{center}
       \includegraphics[width=#2\textwidth, keepaspectratio, #3]{#1}
       \end{center}
       \caption{#5}
       \label{#4}
       \end{figure}
}

\newcommand{\apic}[5]{
       \begin{figure}[H]
       \begin{center}
       \includegraphics[width=#2\textwidth, keepaspectratio, #3]{#1}
       \end{center}
       \caption{#5}
       \label{#4}
       \end{figure}
}

\newcommand{\sapic}[5]{
       \begin{figure}[P]
       \begin{center}
       \includegraphics[width=#2\textwidth, keepaspectratio, #3]{#1}
       \end{center}
       \caption{#5}
       \label{#4}
       \end{figure}
}

\newcommand{\picwrap}[9]{
       \begin{wrapfigure}{#5}{#6}
       \vspace{#7}
       \begin{center}
       \includegraphics[width=#2\textwidth, keepaspectratio, #3]{#1}
       \end{center}
       \caption{#9}
       \label{#4}
       \vspace{#8}
       \end{wrapfigure}
}

\newcommand{\baseT}[2]{\mbox{$#1\times10^{#2}$}}
\newcommand{\baseTsolo}[1]{$10^{#1}$}
\newcommand{\THL}{$T_{\nicefrac{1}{2}}$}

\newcommand{\UBI}{$\rm cts/(kg \cdot yr \cdot keV)$}

\newcommand{\Uflux}{$\rm m^{-2} s^{-1}$}
\newcommand{\Ucpd}{$\rm cts/(kg \cdot d)$}
\newcommand{\Uexpo}{$\rm kg \cdot d$}

\newcommand{\Qbb}{$Q_{\beta\beta}$}

\newcommand{\validate}{\textcolor{blue}{\textit{(validate!!!)}}}

\newcommand{\improve}{\textcolor{blue}{\textit{(improve!!!)}}}

\newcommand{\missing}{\textcolor{red}{\textbf{...!!!...} }}

\newcommand{\quanta}{\textcolor{red}{\textit{(quantitativ?) }}}

\newcommand{\misscite}{\textcolor{red}{[citation!!!]}}

\newcommand{\missref}{\textcolor{red}{[reference!!!]}\ }

\newcommand{\PC}{$N_{\rm peak}$}
\newcommand{\BIC}{$N_{\rm BI}$}
\newcommand{\PAPR}{$R_{\rm p/>p}$}

\newcommand{\PCR}{$R_{\rm peak}$}


\newcommand{\gline}{$\gamma$-line}
\newcommand{\glines}{$\gamma$-lines}

\newcommand{\gray}{$\gamma$-ray}
\newcommand{\grays}{$\gamma$-rays}

\newcommand{\bray}{$\beta$-ray}
\newcommand{\brays}{$\beta$-rays}

\newcommand{\betas}{$\beta$'s}


\newcommand{\tab}{\textcolor{brown}{Table~}}
\newcommand{\eq}{\textcolor{brown}{Eq.~}}
\newcommand{\fig}{\textcolor{brown}{Fig.~}}
\renewcommand{\sec}{\textcolor{brown}{Sec.~}}
\newcommand{\chap}{\textcolor{brown}{Chap.~}}

 \newcommand{\fn}{\iffalse \fi} 
 \newcommand{\tx}{\iffalse \fi} 
 \newcommand{\txe}{\iffalse \fi} 
 \newcommand{\sr}{\iffalse \fi} 
\today

%
%
%
\title{ An improved half-life limit of the double beta decay of $^{94}$Zr into the excited state of $^{94}$Mo}


\author{N. Dokania$^{a,b}$, D. Degering$^c$, B. Lehnert$^d$,  V. Nanal$^a$, K. Zuber$^e$}
\address{
$^a$ Department of Nuclear and Atomic Physics, Tata Institute of Fundamental Research, Mumbai 400 005, India\\
$^b$ Present address: Department of Physics and Astronomy, State University of New York at Stony Brook, NY 11794-3800, USA\\
$^c$ VKTA - Strahlenschutz, Analytik \& Entsorgung Rossendorf e.V., 01328 Dresden, Germany\\
$^d$ Physics Department, Carleton University, 1125 Colonel By Drive, K1S 5B6 Ottawa, Canada\\
$^e$ Institut f\"ur Kern- und Teilchenphysik, Technische Universit\"at Dresden,\\
Zellescher Weg 19, 01069 Dresden, Germany\\
}

\begin{abstract}
A search for the double beta decay transition of $^{94}$Zr into the first excited state of $^{94}$Mo has been performed at the Felsenkeller underground laboratory in Dresden, Germany. A 341.1~g zirconium sample with natural isotopic composition has been measured for 43.9~d in an ultra low background gamma spectroscopy setup. No signal has been observed and a new best lower half-life limit is set as \baseT{5.2}{19}~yr (90\% CI). This limit is valid for the \bb{0} and \bb{2} decay into excited states of $^{94}$Mo but cannot distinguish between the two modes. Existing limits are improved by 50\%.
\end{abstract}


\noindent{\it Keywords}: Double-beta decay, Half-lives

\submitto{\JPG}

\maketitle

\section{Introduction}
\label{intro}

In recent years the field of neutrino physics has made tremendous progress by 
establishing neutrino flavor oscillations in the lepton sector. This is explained by neutrino oscillations requiring a non zero neutrino mass; however, no absolute mass scale can be fixed with experiments studying the oscillatory behavior. New experiments like KATRIN, ECHo and Holmes 
 are starting to explore the mass region below 1 eV in the near future \cite{betaReview}. Two alternative methods
to learn about neutrino masses are cosmology, measuring the sum of all neutrino masses, and 
neutrino-less double beta (\bb{0}) decay.

Investigating neutrinoless double beta decay helps to fix the absolute neutrino mass scale and even allows to study total lepton number violation. 
The \bb{0} decay is the golden channel to distinguish whether neutrinos are Majorana or Dirac particles. This second order weak decay violates lepton number by two units and thus is not allowed in the Standard Model (SM):
\be
(Z,A) \ra (Z+2,A) + 2 e^-  \quad \, . 
\ee

Furthermore, a match of helicities of the intermediate neutrino states is necessary which is done in the easiest way by introducing a neutrino mass. This mass is linked with the experimentally observable half-life via
\begin{equation}
  \label{eq:1}
 \left(T_{1/2}^{0 \nu}\right)^{-1} = G^{0 \nu} \left| M^{0\nu}\right|^2 \left(\frac{\ema}{m_e}\right)^2 \, ,
\end{equation}

where \ema\ is the effective Majorana neutrino mass, given by $\ema = \left| \sum_{i}U_{ei}^2m_i\right|$ with $U_{ei}$ the corresponding  element in the 
leptonic PMNS mixing matrix. $G^{0 \nu}$ is a phase space factor containing the $\ln{2}$ for half-life conversion by convention and $M^{0\nu}$ describes the nuclear transition matrix element. The calculation of $M^{0\nu}$ is based on nuclear model assumptions and is currently the larges uncertainty to convert a measured \bb{0} half-life into $\ema$.
In addition, the SM process of neutrino accompanied double beta decay,
\be
(Z,A) \ra (Z+2,A) + 2 e^- + 2 \bnel \quad (\zbb) \,
\ee
can be investigated and has been observed in about a dozen isotopes. This mode is important
for  understanding the corresponding nuclear matrix element $\left| M^{2\nu}\right|$ which is
solely based on Gamow-Teller transitions. 
While $\left| M^{2\nu}\right|$ is numerically and conceptually different to $\left| M^{0\nu}\right|$ ,it is based on the same nuclear models and can be used to validate nuclear model parameters such as the quenching of the axial vector coupling $g_A$ in nuclear matter \cite{Barea15}.
Depending on the Q-value, this decay can also occur
into one or several excited states of the daughter nucleus. This process is suppressed in 
phase space due to the lower available Q-value, but has the experimental benefit of observing
the associated \grays. The observation of excited state transitions in combination with the ground state transition can additionally constrain nuclear models when many of the model assumptions are the same in a given isotope \cite{Lehnert15}.\\

An interesting element with two double beta decay isotopes is zirconium. Natural zirconium contains two double beta isotopes: \nuc{Zr}{96} with $(2.80\pm 0.09)\%$ isotopic abundance and \nuc{Zr}{94} with $(17.38\pm0.28)\%$ isotopic abundance. The decay of \nuc{Zr}{96} does not exclusively occur via a double beta decay, as for most other double beta decay isotopes, but is also possible via two consecutive beta decays.  However, the beta decay of \nuc{Zr}{96} is fourth forbidden with a $0^+\rightarrow4^+$ transition to the excited state of 146.1~keV and even sixth forbidden with $0^+\rightarrow6^+$ to the ground state and thus so slow that it can compete with the double beta decay process.\\

\begin{table}[htbp]
\begin{center}
\begin{tabular}{lll}
\hline
Decay mode & $T_{1/2} $ [yr] &  Reference\\
\hline
\nuc{Zr}{94} $0/2\nu\beta\beta\ \rm 2^{+}_1 (871.1~keV)$            &  Ex: $> 3.4 \times 10^{19}$ (\unit[90]{\%} CL) &  2017 \cite{Dokania17}\\
\hline
\hline
\nuc{Zr}{96} $\beta^{-}$                                      & Ex: $> 2.4 \times 10^{19}$ (\unit[90]{\%} CL) & 2016 \cite{Finch16}\\
\nuc{Zr}{96} $\beta^{-}$                                      & Th: $=2.4 \times 10^{20}$ & 2007 \cite{Heiskanen07}\\
\hline
\nuc{Zr}{96} $2\nu\beta\beta\ \rm g.s.$              & Ex: $=(2.35\pm0.14_{\rm stat}\pm0.16_{\rm syst}) \times 10^{19}$ & 2010 \cite{Nemo10}\\
\nuc{Zr}{96} $0/2\nu\beta\beta\ \rm 2^{+}_1 (778.2~keV)$      & Ex: $> 4.1 \times 10^{19}$ (\unit[90]{\%} CL)  & 1994 \cite{Arpesella94}\\
\nuc{Zr}{96} $0/2\nu\beta\beta\ \rm 0^{+}_1 (1148.1~keV)$     & Ex: $> 3.1 \times 10^{20}$ (\unit[90]{\%} CL) &   2015 \cite{Finch15}\\
\nuc{Zr}{96} $0/2\nu\beta\beta\ \rm 0^{+}_2 (1330.0~keV)$     & Ex: $> 1.4 \times 10^{20}$ (\unit[90]{\%} CL) &   2015 \cite{Finch15}\\
\nuc{Zr}{96} $0/2\nu\beta\beta\ \rm 2^{+}_2 (1497.8~keV)$     & Ex: $> 1.0 \times 10^{20}$ (\unit[90]{\%} CL) &   2015 \cite{Finch15}\\
\nuc{Zr}{96} $0/2\nu\beta\beta\ \rm 2^{+}_3 (1625.9~keV)$     & Ex: $> 1.2 \times 10^{20}$ (\unit[90]{\%} CL) &   2015 \cite{Finch15}\\
\nuc{Zr}{96} $0/2\nu\beta\beta\ \rm 0^{+}_3 (2622.5~keV)$     & Ex: $> 1.1 \times 10^{20}$ (\unit[90]{\%} CL) &   2015 \cite{Finch15}\\
\hline
\end{tabular}
\medskip
\caption{\label{tab:PrevResults} Currently known half-life limits on various \nuc{Zr}{94} and \nuc{Zr}{96} decay modes. 
The excited state energies for the transitions are given in brackets.
}
\end{center}
\end{table}

\begin{figure}
\centering
\includegraphics[width=0.6\textwidth]{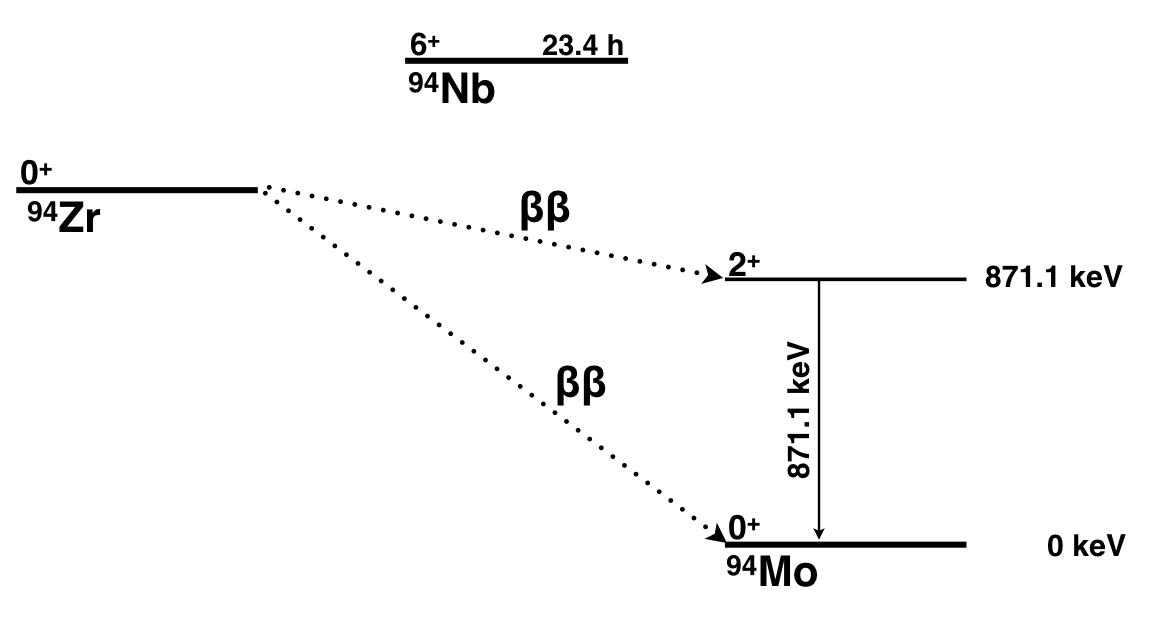}
\caption{%
Decay scheme of \nuc{Zr}{94}.
\label{fig:decaySchemeZr94}}
\end{figure}

Double beta decays of \nuc{Zr}{94} and \nuc{Zr}{96} and beta decays of \nuc{Zr}{96} have been studied in the past. The best current upper half-life limits are listed in \tab \ref{tab:PrevResults}. The measurement in \cite{Finch15} used a zirconium sample isotopically enriched in \nuc{Zr}{96} which increases the sensitivity to study this isotope, but also decreases the sensitivity to study \nuc{Zr}{94}. In this paper we use a 341.1~g natural zirconium sample and focus on the search for the double beta decay of \nuc{Zr}{94} into the first excited state of \nuc{Mo}{94}. This state relaxes under the emission of a single \gray\ of 871.1~keV as shown in the decay scheme in \fig \ref{fig:decaySchemeZr94}. The search is performed with a high purity germanium (HPGe) detector and cannot distinguish between the \bb{0} and \bb{2} mode of the decay. The obtained limit is valid for both cases.

%


\section{Experimental Setup, Sample and Data Taking}
\label{setup}

The detector setup is placed in the Felsenkeller Underground Laboratory in Dresden, Germany. The laboratory has an overburden of 110~m.\ w.\ e.\ reducing the muon flux to  \unit[\baseT{0.6}{-3}]{cm$^{-2}$s$^{-1}$} \cite{Niese98,FelixDPG}. The detector itself is an ultra low background HPGe detector of \unit[90]{\%} relative efficiency, routinely used for low background gamma spectroscopic measurements. The sample compartment is surrounded by a \unit[5]{cm} copper shielding embedded in another shielding of \unit[15]{cm} of clean lead. The inner \unit[5]{cm} of the lead shielding have a specific activity of \unit[($2.7 \pm 0.6$)]{Bq/kg}  $^{210}$Pb while the outer \unit[10]{cm} have \unit[($33 \pm 0.4$)]{Bq/kg}. The spectrometer is located in a measuring chamber with additional shielding. The sample compartment is constantly held in a nitrogen atmosphere to avoid radon and its daughters \cite{deg09,degering08}.\\

\begin{figure}
\centering
\includegraphics[width=0.95\textwidth]{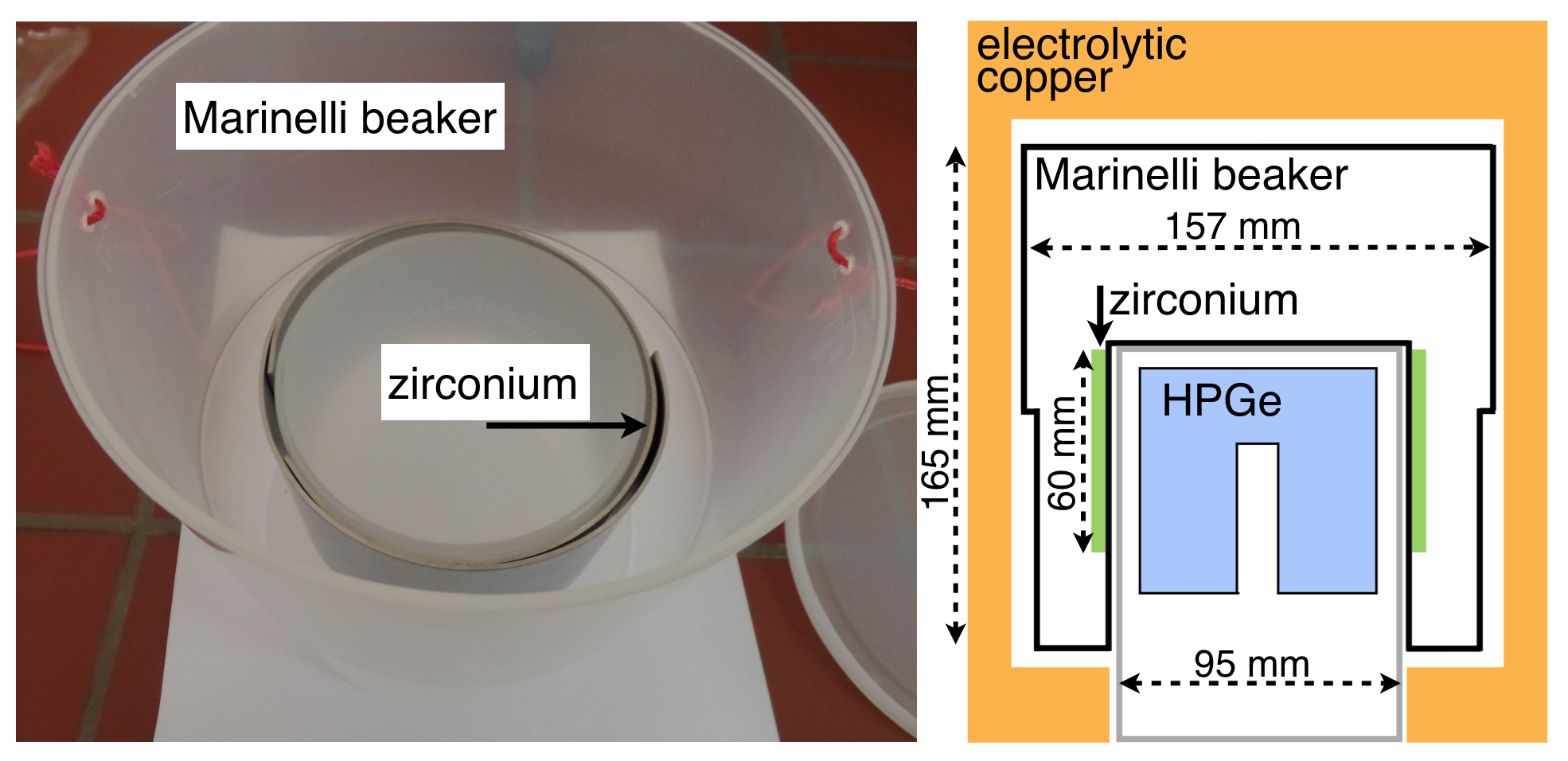}
\caption{%
Picture and schematic of the 341.1~g natural zirconium sample in the measurement setup. The sample consists of two 300~mm x 60~mm x 1.5~mm thick zirconium plates which are wrapped around the inner side of a Marinelli beaker.
\label{fig:Sample}}
\end{figure}

The sample of 341.1~g natural zirconium is in the shape of two plates with 1.5~mm thickness as shown in \fig \ref{fig:Sample}. The plates are wrapped around the inner sides of a standard \unit[1.5]{l} Marinelli beaker facing the lateral side of the HPGe detector.
The detector system and sample arrangement were implemented in the code framework MaGe \cite{Boswell:hc} developed for simulations of HPGe detectors and tuned to low energy particle propagation required for gamma spectrometry measurements. \\

The data were collected with a 8192 channel MCA from ORTEC recording energies up to \unit[2.8]{MeV} for three independent measurement periods of M1 = 10.6~d, M2 = 19.8~d and M3 = 13.5~d. The total measuring time of 43.9~d translates into a total exposure of 15.0~kg x d or an isotopic exposure of 2.60~kg x d for \nuc{Zr}{94} and 0.42~kg x d for \nuc{Zr}{96}. This compares to 3.42~kg x d of \nuc{Zr}{96} exposure for the 91.4\% enriched sample in the previous measurement of \cite{Finch15} and to 14.7~kg $\times$ d \nuc{Zr}{94} exposure for the natural Zr sample in \cite{Dokania17}.\\

\begin{figure}
\centering
\includegraphics[width=0.75\textwidth]{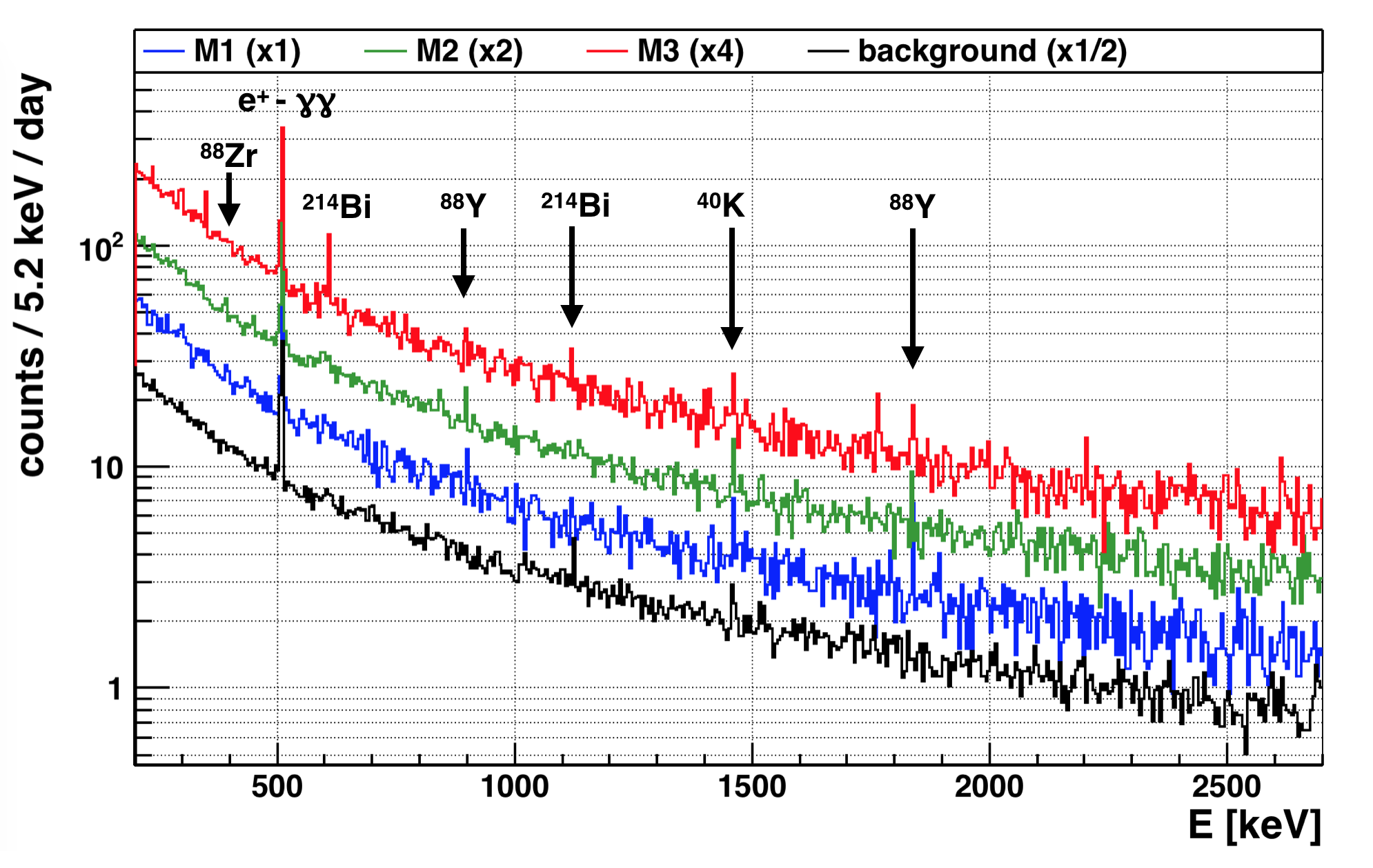}
\caption{%
Individual zirconium sample spectra of all three measurements and a background spectrum normalized to counts per day and 5.2~keV bin. The y-axis of M2, M3 and the background spectrum are scaled by a factor of 2, 4 and 1/2, respectively for visibility as indicated in the legend.
\label{fig:SumSpec}}
\end{figure}

The three measurement periods as well as a background spectrum for comparison are shown in \fig \ref{fig:SumSpec}. 
Apart from natural decay chain isotopes, two clear \glines\ at 898.9~keV and 1836.1~keV are identified from \nuc{Y}{88}  and one \gline\ at 392.9~keV from \nuc{Zr}{88}. 
This is due to the fact that the sample has spent a significant time on surface and was shipped from India to Germany by plane.  This caused some activation of the material. As \nuc{Zr}{90} has the largest abundance of all Zr-isotopes (51.45\%) the production of \nuc{Y}{88} can occur via (n,3n) reaction with the subsequent electron capture of \nuc{Zr}{88}
or (n,t) reactions. Other potential activation candidates have short half-lives and have decayed away. \\

The 898.9~keV (93.7\%) \gline\ and 1836.1~keV (99.2\%) \gline\ from \nuc{Y}{88} (\THL=106.6~d) 
as well as the 392.9~keV (97.3\%) \gline\ from \nuc{Zr}{88} (T$_{1/2}$=83.4~d)
are only identified in the first two of the three measurements. 
The measurement details and peak count rates for the \nuc{Y}{88} and \nuc{Zr}{88} \glines\ are shown in \tab \ref{tab:Y88}. The first measurement M1 and the second measurement M2 are 25~d apart which corresponds to a 20\% reduced peak count rate given the half-life of \nuc{Zr}{88}. A reduction of the count rate is indeed observed for \nuc{Zr}{88} and \nuc{Y}{88}, however, the difference is not significant within counting uncertainties. The third measurement M3 was performed 230~d after M1 and thus the \nuc{Zr}{88} peak count rates are expected to be reduced to 15\% compared to M1. In the data of M3 the \glines\ do not show significant counts above background which is consistent with expectation of the half-lives of \nuc{Zr}{88} and \nuc{Y}{88}.\\

%
%
%



\begin{table}[htbp]
\begin{center}
\begin{tabular}{ccccccc}
\hline
dataset & start date & $\Delta$T since M1 & livetime  & 898.9~keV &  1836.1~keV & 392.9~keV\\
	& [dd/mm/yy]  & [d]  & [d]  & [cts/d]  & [cts/d]  & [cts/d]\\
\hline
M1 & 14/07/16    &  0     & 10.58 & $5.0\pm1.0$  & $3.9\pm0.8$  & $5.1\pm1.3$\\
M2 & 08/08/16    &  25   & 19.77  & $4.2\pm0.7$  & $2.8\pm0.5$  & $3.1\pm0.9$\\
M3 & 03/01/17    &  230 & 13.54  & $0.8\pm0.9$  & $-0.2\pm0.5$ & $0.6\pm0.9$\\
\hline
\end{tabular}
\medskip
\caption{\label{tab:Y88} Peak count rates for \nuc{Y}{88} and \nuc{Zr}{88} background for the three different measurement periods.
}
\end{center}
\end{table}

\section{Analysis and Results}
\label{analysis}

The search for the 871.1~keV \gline\ from the double beta decay of \nuc{Zr}{94} into the first excited state of \nuc{Mo}{94} is performed with a fit to the three individual spectra of the measurements. The detector was stable over the $\approx$250~d spanning three measurements. However, energy calibrations have been performed in-between.

\begin{figure}
\centering
\includegraphics[width=0.99\textwidth]{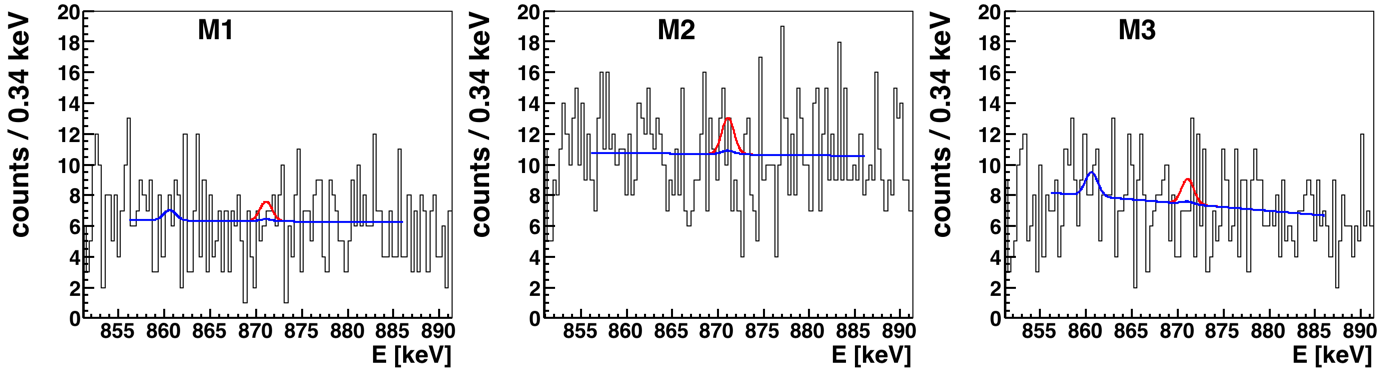}
\caption{%
Spectral fit of the \nuc{Zr}{94} double beta decay \gline\ at 871.1~keV and the background \gline\ of \nuc{Tl}{208} at 860.6~keV. The solid blue line shows the best fit and the solid red line shows the signal peak at 90\% CI. 
\label{fig:pdf_ROI_Composition_Zr94DBD_2p1}}
\end{figure}

The fit is performed with a single half-life parameter connecting the three individual datasets $d$. The fit range is chosen as $\pm 15$~keV around the 871.1~keV \gline\ in order to model the background with a linear function in the vicinity of the peak (see \fig \ref{fig:pdf_ROI_Composition_Zr94DBD_2p1}). The \nuc{Tl}{208} 860.6~keV background \gline\ with 12.5\%  emission probability (4.49\% within the \nuc{Th}{232} chain) is in this region and included in the fit.

The signal counts $s_{d}$ of the 871.1~keV \gline\ in each dataset are connected with the half-life $T_{1/2}$ of the decay mode as
\begin{eqnarray}
\label{eq:PdHLtoCounts}
s_{d} =
\ln{2} \cdot  \frac{1}{T_{1/2}} \cdot \epsilon \cdot N_A \cdot T_d \cdot m \cdot f  \cdot \frac{1}{M}\ ,
\end{eqnarray}

where $\epsilon$ is the full energy detection efficiency,
$N_A$ is the Avogadro constant,
$T_d$ is the live-time of dataset $d$, 
$m$ is the mass of the zirconium-sample, 
$f$ is the natural isotopic abundance of \nuc{Zr}{94} 
and $M$ the molar mass of natural zirconium. 
%
The Bayesian Analysis Toolkit (BAT) \cite{Caldwell:2009kh} is used to perform a maximum posterior fit combining all three datasets. The likelihood $\mathcal{L}$ is defined as the product of the Poisson probabilities over each bin $i$ in dataset $d$ for observing $n_{d,i}$ events while expecting $\lambda_{d,i}$ events:
\begin{eqnarray}
\mathcal{L}(\mathbf{p}|\mathbf{n}) =
\prod \limits_d \prod  \limits_i \frac{\lambda_{d,i}(\mathbf{p})^{n_{d,i}}}{n_{d,i}!} e^{-\lambda_{d,i}(\mathbf{p})}\ ,
\end{eqnarray}

where \textbf{n} denotes the data and \textbf{p} the set of floating parameters.
$\lambda_{d,i}$ is taken as the integral of the extended p.d.f.\ $P_{d,r}$ in this bin
\begin{eqnarray}
\lambda_{d,i}(\mathbf{p}) &=&
 \int_{\Delta E_{d,i}} P_{d}(E|\mathbf{p}) dE\ , \label{eq:Ta_expcounts}
\end{eqnarray}

where $\Delta E_{d,i}$ is the bin width in each dataset $d$. 
The counts in the fit region are expected from three different types of contributions which are used to construct $P_{d}$:
(1) a linear background, (2) the Gaussian signal peak and (3) the Gaussian background peak. The full expression of $P_{d}$ is written as:
\begin{eqnarray}
P_{d}(E|\mathbf{p}) &&=
 B_{d} + C_{d}\left( E-E_0 \right) \label{eq:Ta_pdf}\\[2mm]
&&+  \frac{s_{d}}{\sqrt{2\pi}\sigma_{d}} 
\cdot \exp{\left(-\frac{(E-871.1~{\rm keV})^2}{2\sigma_{d}^2}\right)}\nonumber\\[2mm]
&&+  \frac{b_{d}}{\sqrt{2\pi}\sigma_{d}} 
\cdot \exp{\left(-\frac{(E-860.6~{\rm keV})^2}{2\sigma_{d}^2}\right)}.\nonumber
\end{eqnarray}

The first row is describing the linear background with the two parameters $B_{d}$\ and $C_{d}$.
The second line is describing the signal peak with the energy resolution $\sigma_{d}$ and the \gline\ energy as the mean of the Gaussian.
The third line is describing the background \gline\ with the strength of the peak $b_{d}$. The same p.d.f.\ with different parameter values is used for all three datasets (\fig \ref{fig:pdf_ROI_Composition_Zr94DBD_2p1}).\\

Each free parameter in the fit has a prior associated. The prior for the inverse half-life $(T_{1/2})^{-1}$ is flat. Priors for energy resolution, peak position and detection efficiencies are Gaussian, centred around the mean values of these parameters. The width of these Gaussian is the uncertainty of the parameters. 
This naturally includes the systematic uncertainty into the fit result.
The uncertainty of the peak positions are set to 0.1~keV.    
The energy scale and resolution is routinely determined using reference point sources including \nuc{Am}{241}, \nuc{Cs}{137} and \nuc{Co}{60}. The main \glines\ of these radionuclides are fitted by a Gaussian distribution and the energy resolution is interpolated by a quadratic function. A resolution of $\sigma=0.67$~keV was determined at \unit{871.1}{keV} with an estimated uncertainty of 10\%.
The full energy peak detection efficiencies determined with MC simulations is 4.1\% at 871.1~keV with an estimated uncertainty of 10\%. Uncertainties on the measured sample mass, the measuring time and the isotopic abundance enter the fit in the same way as the detection efficiency but are negligible in comparison.  \\

The posterior probability distribution is calculated from the likelihood and prior probabilities with BAT. The maximum of the posterior is the best fit. The posterior is marginalized for $(T_{1/2})^{-1}$ and used to extract the half-life limit with the \unit[90]{\%} quantile. This results in \unit[90]{\%} credibility limits on the half-life (red line in \fig \ref{fig:pdf_ROI_Composition_Zr94DBD_2p1}).  
The 90\% quantile of the inverse half-life distribution is \baseT{1.92}{-20}~yr$^{-1}$ which translates into a 90\% C.I.\ lower half-life limit of \baseT{5.2}{19}~yr. Fixing the systematic uncertainties in the fit improves the limit by 0.2\%. Varying the fit range from $\pm15$~keV to $\pm13$~keV and $\pm17$~keV reduces the half-life limit by 3\% in both cases. \\

Parameter $B_d$ in \eq \ref{eq:Ta_pdf} is used to extract the background level from the fit as $1.72 \pm 0.08$~cts/keV/d for M1, $1.55 \pm 0.05$~cts/keV/d for M2 and $1.59 \pm 0.06$~cts/keV/d for M3. This is one order of magnitude lower background than for the search in \cite{Dokania17} with about 13.3~cts/keV/d. 
A detailed background model is not available for the detector setup. However, the \gray\ background is rather small (\fig \ref{fig:SumSpec}) with $3.2\pm0.8$~cts/d for the 238.6~keV \gline\ from \nuc{Pb}{212} and $1.9\pm0.4$~cts/d for the 1460.8~keV \gline\ from \nuc{K}{40} mainly coming from the detector materials. The spectrum is dominated by a continuous component from direct muons and muon induced neutron interactions.\\

An additional analysis of \nuc{Zr}{96} was performed but did not yield better limits than reported in \cite{Arpesella94} and \cite{Finch15}. Applying the same technique as described for \nuc{Zr}{94}, the transition into the first excited $2^+$ with a de-excitation \gline\ at 768.4~keV was investigated. No signal was found and a 90\% CI lower half-life limit of \baseT{1.8}{19}~yr was obtained. For the $0^+$ transition a combined fit of the 768.4~keV and 369.8~keV \gline\ was performed similar to the analysis in \cite{Pd}. The obtained lower half-life limit is \baseT{2.0}{19}~yr.

\section{Conclusions}
\label{conclusions}

The half-life of the double beta decay transition of $^{94}$Zr  into the first excited state of $^{94}$Mo has been investigated with a low background gamma spectroscopy setup at the Felsenkeller underground laboratory in Dresden, Germany. No signal has been observed and a new best lower half-life limit is set as \baseT{5.2}{19}~yr (90\% CI). This is a 50\% improvement compared to the previous best limit in \cite{Dokania17}. The improvement could be achieved with a 6 times lower exposure compared to the previous best limit due to a detector setup with about one order of magnitude lower background. The sensitivity for the single and double beta decay of \nuc{Zr}{96} could not be improved compared to previous searches with an enriched sample \cite{Finch15}.\\

The sensitivity of this search could be improved with a longer measurement time, larger sample size and using isotopically enriched material. Typically, larger sample masses would decrease the detection efficiency of de-excitation \grays\ due to self-absorption in the sample; however, the rather small mass of  341.1~g in this search suggests ample room for improvement, especially in connection with a dedicated optimization of the sample - detector arrangement based on MC simulations. 
The already low background environment of the detector is dominated by muons which could be reduced with an active muon veto for the setup. However, this is limited due to muon induced neutrons which ultimately require a deeper underground location for significant background reduction.

%

\end{document}